# STEADY COEXISTENCE OF THE SUBJECTS OF THE MARKET REPRESENTING THE PRIVATE AND STATE CAPITAL

Viktor I. Shapovalov

The Volgograd Branch of Moscow Humanitarian-Economics Institute, Volgograd, Russia.
shavi@rol.ru

The sustainability conditions for the market participants with a different ownership model were also determined. It was revealed, that the nonlinear form of the equations describing the market behavior with the prevailing private capital, predetermines the development of such a market according to the subharmonic cascade scenario. The latter is presumably the reason of the periodically arising economic crises.

In the present work we shall determine control parameters and those intervals of their values at which steady coexistence of the subjects of the market representing the private and state capital is possible [1].

Let $X$ and $Y$ are identical or interchangeable goods. The goods $X$ are realized by private enterprise; $Y$ – by state one. In most cases expected volume of the next sale can be predicted by results of previous one. Last thing is characteristic for Markov processes; therefore analysis of stability is convenient for carrying out by a method of dot reflections. Let's enter designations: $X_{n+1}$ (or $Y_{n+1}$) – expected volume of the next sale; $X_n$ (or $Y_n$) – volume of last sale. The units of measurements have no basic importance.

In the case of the private seller expected volume of the next sale $X_{n+1}$ is proportional to volume $J_1$, presented to sale, and demand $J_2$ on the goods $X$: $X_{n+1} = \varepsilon J_1 J_2$, where $\varepsilon$ – factor of proportionality. As a rule, the private businessman while defining quantity presented for sale goods is guided by sales quantity in previous times, i.e. volume $J_1$ should be proportional to $X_n$. Besides we believe that a major factor determining average (in examined region) demand for this or that goods is the income of average (in the same region) buyer. In our designations it looks so: the demand $J_2$ for the goods $X$ is proportional to the income $C_0$ of the average buyer. From this demand it is necessary to subtract that its part $J_3$, which was satisfied with purchase $Y$ instead of $X$. In turn $J_3$ is proportional to the price $\beta_x$ for the goods $X$ (it means, that the more price $\beta_x$ is – the more sales of $Y$ instead of $X$ are) and number of crossings of the goods $X$ and $Y$. By crossing of these goods is understood their joint presence at the market, at which the buyer has opportunity to make a choice for the benefit of one of them (for example, they are on the next counters or in next shops).

In result we have:

$$X_{n+1} = X_n(\alpha C_0 - \mu \beta_x X_n Y_n) \tag{1}$$

– dot reflection describing process of trade of the private seller. Here: $\alpha$ and $\mu$ – factors of proportionality.

In case of the state seller volume presented to sale, $J_1$ is determined by state needs and is constant value $A$ in the sense that it does not depend on number of iterations of dot reflection in the designations, accepted by us. Other proportions do not change. Therefore for the state seller

is received the following reflection:

$$Y_{n+1} = A(\alpha C_0 - \mu \beta_y X_n Y_n), \qquad (2)$$

where $\beta_y$ – price of the goods Y.

We take advantage a technique of research of dot reflection. In result we find stationary volumes of sales of each participant of the market and also stationary relation of the prices on the specified goods:

$$X^* = \frac{C-1}{\delta_x Y^*}; \quad Y^* = \frac{C}{(1/A) + \delta_y X^*}; \quad \frac{\delta_y}{\delta_x} = \frac{C - Y^*/A}{C-1},$$

where the reduced designations are entered: $C = \alpha C_0$ and $\delta_i = \mu \beta_i; i = x, y$.

The stationary decisions we shall check up on stability.

At first we shall define a condition of stability for $X^*$. For this purpose we ought to be concentrated on features of behavior of each separate participant of the market. In this connection we shall make indignation of the stationary decision $X^*$, leaving $Y^*$ fixed. Let's remind, that in the theory of dot reflections the stationary decision (the so-called motionless point) is steady, if the following condition is satisfied

$$\left| \left( \frac{\partial f_x}{\partial X_n} \right)_{X^*} \right| < 1,$$

where $f_x = F_x(X_n, Y^*)$ – the right part of reflection (1), in which $Y_n$ is replaced by the fixed value of $Y^*$. In result we find:

$$1 < \alpha C_0 < 3 \qquad (3)$$

– a condition of steady activity of the participant of the market with a private pattern of ownership.

Similarly we shall define a condition of stability for $Y^*$ (in the right part of reflection (2) $X_n$ is replaced with the fixed value $X^*$):

$$0 < \alpha C_0 < \frac{1}{1 - 0{,}5 \beta_x / \beta_y} \qquad (4)$$

–a condition of steady activity of the participant of the market with a state pattern of ownership.

If we solve this inequality together with an inequality (3), then we shall find a condition of steady coexistence in the market of the representatives of private and state patterns of ownership:

$$1 < \alpha C_0 < \frac{1}{1 - 0{,}5 \beta_x / \beta_y} \le 3. \qquad (5)$$

From an extreme right part of this expression it is possible to receive a ratio for the prices established by the participants of the market:

$$\beta_x \le \frac{4}{3} \beta_y. \qquad (6)$$

Let's make the obvious assumption: the state prices are less mobile, than prices of the representatives of the private capital. Then from an inequality (6) follows, that the stability of the market will not be disturbed (i.e. volume of sales of the participants of the market will make steadily value *X\** and *Y\**) even in the event when the private seller begins to sell his goods little bit more expensively then state one in the specified limit: the average price $\beta_x$ can be on 30% more then average price $\beta_y$.

The given conclusion has the important consequence: the system properties of the private capital' representatives give them an opportunity to receive the greater income in comparison with the state subjects within other equal conditions. In the present research the other equal conditions were taken into account by introduction in the equations (1) and (2) just the same factors $\alpha$ and $\mu$.

The same conclusion is indirectly proved by analysis of conditions of steadiness (3) and (4). It is possible to assume on the basis of the left part of a condition (3) that for the market participant with a private pattern of ownership the minimal volume of the next sale should not come to zero. Indeed, it is obvious from the equation (1), that under favorable conditions, when the competition is small (i.e. $\mu\beta_x X_n Y_n \to 0$), the low border of a condition of stability (3) requires, that $X_{n+1} \cong X_n$. On the contrary, for the state market' participant the low border of a condition (4) supposes zero volume of the next sale. In other words, the system features of the private businessman compel him to work economically more effectively, than state one.

Now we address to the right part of an inequality (3): with growth of the income $C_0$ of the average buyer the top limit is broken, and the private businessman can lose steadiness. To understand the reason of it, we shall make the replacement of variables

$$Z_n = X_n \frac{\mu\beta_x Y^*}{\alpha C_0}; \quad Z_{n+1} = X_{n+1} \frac{\mu\beta_x Y^*}{\alpha C_0};$$

and we shall enter a designation $4\gamma = \alpha C_0$. In result the reflection (1) will be transformed to a form

$$Z_{n+1} = 4\gamma Z_n (1 - Z_n). \tag{7}$$

This expression is a well-known nonlinear reflection admitting the chaotic system development script by subharmonic cascade. As it is known [3], the stationary state of reflection (7) loses steadiness at $\gamma > 0,75$ (first doubling of quantity of steady states). Whence we receive: $\alpha C_0 > 3$, that in turn corresponds to infringement of the right part of a condition (3). Hence, the market with the prevailing private capital develops under the script of the subharmonic cascade [1].

Last means, that at the further increase $C_0$ it is necessary to expect occurrence of turns of quantity doubling of steady *X\**. This process inevitably should be finished by a chaotic state (for this reason the subharmonic cascade has received the name "the script of chaos"). Apparently, in practice, last finds the reflection as crises of prevailing private capital economics.

Let's notice, that the reflection (2), written down for the state ownership pattern representatives is linear by $Y_n$ and cannot be transformed to a nonlinear form (7). Probably, it explains the absence of periodic crises in economy controlled by the state.

Let's get a brief result on the considered model.

1. The inequality (5) represents a condition, at which the steady stationary coexistence of the subjects of the market with private and state patterns of ownership is possible. From this condition, and also from conditions (3) and (4) follows, that for system "market" as the basic control parameters acts the income of the average buyer and relation of the prices established by the participants of the market. By changing the specified parameters, we can break any of these conditions. In result system "market" destabilizes at once.

2. It was revealed, that the nonlinear form of the equations describing the market behavior with the prevailing private capital, predetermines the development of such a market according to the subharmonic cascade scenario. The latter is presumably the reason of the periodically arising economic crises.